\documentclass[preprint,12pt]{elsarticle}
\usepackage{graphics}
\usepackage{subfigure}
\usepackage{graphicx}
\usepackage{amsmath}
\usepackage{float}

\usepackage{amssymb}

\journal{PhysicaA}

\begin{document}

\begin{frontmatter}



\title{Directional Transport of Propelled Brownian Particles Confined in a Smooth Corrugated Channel with Colored Noise}

\author{Bing Wang}
\author{Hao Chen}
\author{Yong Wu}
\address{Department of Mechanics and Physics, Anhui University of Science and Technology, Huainan, 232001, P.R.China}

\begin{abstract}
The transport phenomenon(directional movement) of self-propelled Brownian particles moving in a smooth corrugated confined channel is investigated. It is found that large $x$ direction noise intensity should reduce particles directional movement when the angle Gaussian noise intensity is small. The average velocity  has a maximum with increasing $x$ direction noise intensity when angle Gaussian noise intensity is large. $y$ directional noise has negligible effect on $x$ directional movement. Large angle Gaussian noise intensity should weaken the directional transport. Interestingly, the movement direction changes more than once with increasing periodic force amplitude.
\end{abstract}


\end{frontmatter}

\section{\label{label1}Introduction}
Theory of Brownian motion has played a guiding role in the development of statistical physics. This theory provides a link between the microscopic dynamics and the observable macroscopic phenomena such as particles transport. Particles transport in confined periodic structures have been subjects of research for many decades due to their relevance in several physical systems\cite{b.a1,b.a2,b.a3,b.a4,b.a5,b.a6,b.a7,b.a8}. Confined Brownian motion is ubiquitously occurring in the geometrical confinement in nature such as cells, as well as in artificial patterned structures contributed by experimental development. From the natural occurrence to artificial microdevice, zeolites\cite{b.a9}, ion channels\cite{b.a10,b.a11}, artificial nanopores \cite{b.a12,b.a13} as well as micro fluidic devices\cite{b.a14,b.a15, b.a16,b.a17} are explored elaborately for deep understanding.

The control of the properties of tracer transport along confine channels is a key issue for a variety of situations. The transport processes on microscale can exhibit entirely different properties from those encountered in the macroscopic world. Ding \emph{et al}. have investigated the particles transport in a confined channel without external force with oscillatory boundary, and found the average velocity has a maximum with increasing noise intensity, being alike the stochastic resonance phenomenon\cite{b.a18}. Ao \emph{et al}. have investigated the transport diffusivity of  Janus particles in the absence of external biases with reflecting walls, and found the self-diffusion can be controlled by tailoring the compartment geometry\cite{b.a19}. Wu \emph{et al}. have numerically investigated the transport of anisotropic particles in tilted periodic structures, and found the diffusion and mobility of the particles demonstrate distinct behaviors dependence on the shape of the particles\cite{b.a20}. Liu \emph{et al}. have investigated the entropic stochastic resonance when a self-propelled Janus particle moves in a double-cavity container, results indicated the entropic stochastic resonance can survive even if there is no symmetry breaking in any direction\cite{b.a21}. H\"{a}nggi \emph{et al}. proposed a simple scheme for absolute negative mobility of asymmetry particle in a compartmentalized channel or a rough channel\cite{b.a22}. Marino \emph{et al}. showed the rotational Brownian motion of colloidal particles in the over damped limit generates an additional contribution to the ¡±anomalous¡± entropy\cite{b.a23}. Pu \emph{et al}. investigated the reentrant phase separation behavior of active particles with anisotropic Janus interaction, and found that phase separation shows a re-entrance behavior with variation of the Janus interaction strength\cite{b.a24}.

In this paper,we study the transport phenomenon of Brownian particles confined in a two-dimensional smooth channel. The paper is organized as follows: In Section \ref{label2}, the basic model of self-propelled ratchets with a two-dimensional confined channel and colored noise is provided. In Section \ref{label3}, the effects of parameters is investigated by means of simulations.  In Section \ref{label4}, we get the conclusions.

\section{\label{label2}Basic model and methods}
In the present work, we consider the self-propelled particles moving in a two dimensional smooth corrugated channel. The dynamics of Brownian particle is governed by the following Langevin equations\cite{b.a20}
\begin{equation}
\frac{dx}{dt}=v_0\cos\theta+\xi_x(t), \label{Ext}
\end{equation}
\begin{equation}
\frac{dy}{dt}=v_0\sin\theta+\mu f\cos(\omega t)+\xi_y(t), \label{Eyt}
\end{equation}
\begin{equation}
\frac{d\theta}{dt}=\xi_{\theta}(t). \label{Ethetat}
\end{equation}

$(x, y)$ is the position of the particle mass center. The self-propelled speed is $v_0$. $\mu$ represents the mobility. Angle $\theta$ denotes its direction with respect to the channel axis. The particle is driven by an unbiased time periodic force $f\cos(\omega t)$ with amplitude $f$ and angular frequency $\omega$.  $\xi_x$ is the $x$ direction Gaussian colored noise. $\xi_y$ is the $y$ direction Gaussian colored noise. $\xi_{x}$ and $\xi_{y}$ satisfy the following relations
\begin{equation}
\langle\xi_i(t)\rangle=0,(i=x,y)
\end{equation}
\begin{equation}
\langle\xi_i(t)\xi_j(s)\rangle=\delta_{ij}\frac{Q_i}{\tau_i}\exp[-\frac{|t-s|}{\tau_i}],(i=x,y)
\end{equation}
$\langle\cdots\rangle$ denotes an ensemble average over the distribution of the random forces. $Q_i(i=x,y)$ is the noise intensity of $\xi_i(i=x,y)$. $\tau_i(i=x,y)$ is the self-correlation time of $\xi_i(i=x,y)$.

$\xi_{\theta}$ is the self-propelled angle Gaussian white noise, and describes the nonequilibrium angular fluctuation. $\xi_{\theta}$ satisfies the following relations
\begin{equation}
\langle\xi_{\theta}(t)\rangle=0,
\end{equation}
\begin{equation}
\langle\xi{\theta}(t)\xi{\theta}(t')\rangle=Q_{\theta}\delta(t-t'),
\end{equation}
$Q_{\theta}$ is the noise intensity.
\begin{figure}
\center{
\includegraphics[height=8cm,width=10cm]{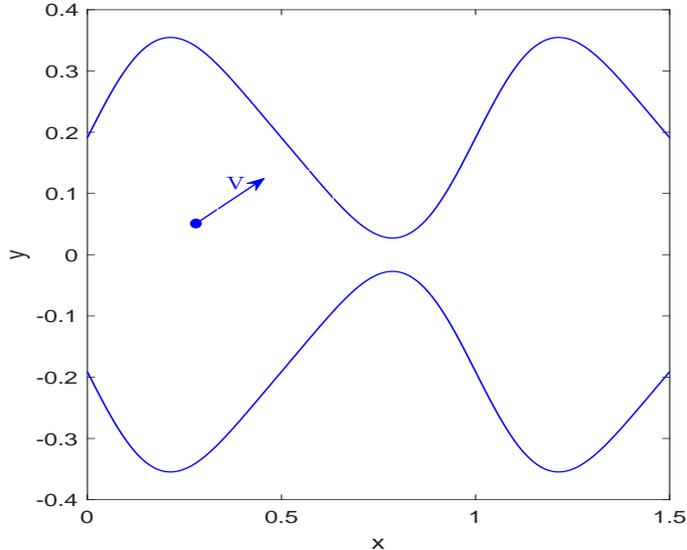}
\caption{Illustrations of the smooth corrugated channel with $a=\frac{1}{2\pi}$, $b=1.2a$, $L=1$, $\Delta=0.5$.}
\label{Channel}}
\end{figure}

The particles are confined in a two dimensional smooth corrugated channel. The channel is periodic in space along the $x$-direction with period $L$, as depicted in Fig.\ref{Channel}. The walls of the cavity have been modelled by the following sinusoidal functions
\begin{align}
W_+(x)=a[\sin(\frac{2\pi x}{L})+\frac{\Delta}{4}\sin(\frac{4\pi x}{L})]+b,\\
W_-(x)=-a[\sin(\frac{2\pi x}{L})+\frac{\Delta}{4}\sin(\frac{4\pi x}{L})]-b,
\end{align}
$a$ controls the slope of walls and $b$ controls the channel width at the bottleneck. $\Delta$ determines the left-right asymmetries and the bottleneck width of the channel. The channel width is $h(x)=W_+(x)-W_-(x)$. The channel midline height is $\frac{W_+(x)-W_-(x)}{2}$.

A central practical question in the theory of Brownian motors is the over all long-time behavior of the particle, and the key quantities of particle transport is the particle velocity $\langle V\rangle$. Because particles along the $y$-direction are confined, we only calculate the $x$-direction average velocity $\langle V\rangle$ based on Eqs.(\ref{Ext},\ref{Eyt},\ref{Ethetat}). $\langle V\rangle$ can be corroborated by Brownian dynamic simulations performed by integration of the Langevin equations using the stochastic Euler algorithm.
 \begin{equation}
\langle V\rangle=\lim_{t\to\infty}\frac{\langle{x(t)-x(t_0)}\rangle}{t-t_0},
\end{equation}
$x(t_0)$ is the position of particles at time $t_0$.

\section{\label{label3}Results and discussion}
In order to give a simple and clear analysis of the system. Eqs.(\ref{Ext},\ref{Eyt},\ref{Ethetat}) is integrated using the Euler algorithm. The total integration time was more than $10^5$ and the integration step time $\Delta t=10^{-4}$. The stochastic averages were obtained as ensemble averages over $10^5$ trajectories. With these parameters, the simulation results are robust and do not depend on the time step, the integration time, and the number of
trajectories.

\begin{figure}
\center{
\includegraphics[height=8cm,width=10cm]{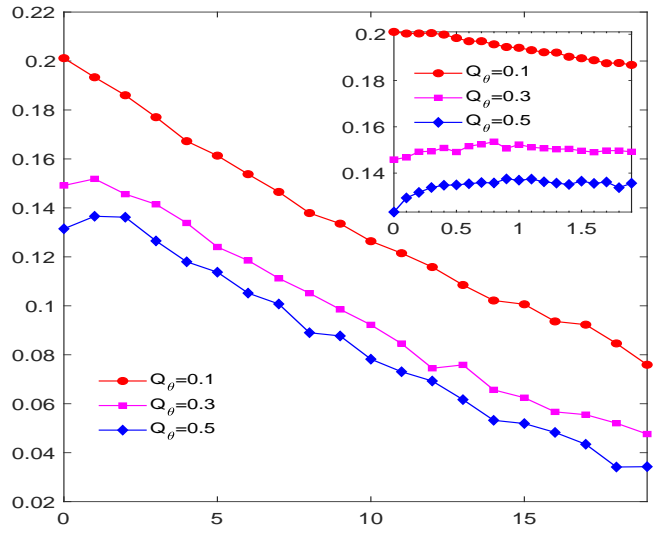}
\caption{The average velocity $\langle V\rangle$ as a function of $x$ direction noise intensity $Q_x$ with different $Q_{\theta}$. The other parameters are $Q_y=2.0$, $v_0=2.0$, $\omega=0.2$, $\Delta=0.2$, $a=1/(2\pi)$, $b=1.2a$, $L=1.0$, $f=3.0$, $\mu=1.0$, $\tau_x=\tau_y=1.0$.}
\label{VQx}}
\end{figure}
The average velocity $\langle V\rangle$ as a function of the $x$ direction noise intensity $Q_x$ with different angle Gaussian noise intensity $Q_{\theta}$ is reported in Fig.\ref{VQx}. It is found that the average velocity $\langle V\rangle$  decreases monotonically with increasing $Q_x$ when $Q_{\theta}=0.1$. But when $Q_{\theta}=0.3$ or $0.5$, $\langle V\rangle$ has a maximum with increasing $Q_x$(the inset figure). So large $x$ direction noise intensity should reduce particles directional movement when the angle Gaussian noise intensity is small($Q_{\theta}=0.1$). With increases of $Q_x$, the directional movement speed first increases, then decreases, when $Q_{\theta}$ is large. In Fig.\ref{VQx}, we also find $\langle V\rangle$ decreases with increasing $Q_{\theta}$.

\begin{figure}
\center{
\includegraphics[height=8cm,width=10cm]{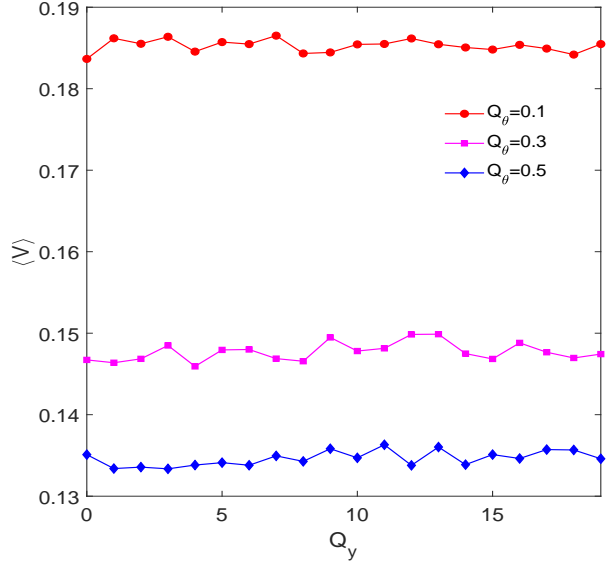}
\caption{The average velocity $\langle V\rangle$ as a function of noise intensity $Q_y$ with different $Q_{\theta}$ . The other parameters are $Q_x=2.0$, $v_0=2.0$, $\omega=0.2$, $\Delta=0.2$, $a=1/(2\pi)$, $b=1.2a$, $L=1.0$, $f=3.0$, $\mu=1.0$, $\tau_x=\tau_y=1.0$.}
\label{VQy}}
\end{figure}
The $x$ directional movement average velocity $\langle V\rangle$ as a function of $y$ directional noise intensity $Q_y$ with different $Q_{\theta}$ is reported in Fig.\ref{VQy}. We find, unlike the effects of $x$ directional noise intensity, the effects of $Q_y$ is simple. The $\langle V\rangle-Q_y$ line tends to be parallel to the horizontal axis.  This means that $y$ directional noise has negligible effect on $x$ directional movement. In Fig.\ref{VQy}, we can also find $\langle V\rangle$ decreases with increasing $Q_{\theta}$.

\begin{figure}
\center{
\includegraphics[height=8cm,width=10cm]{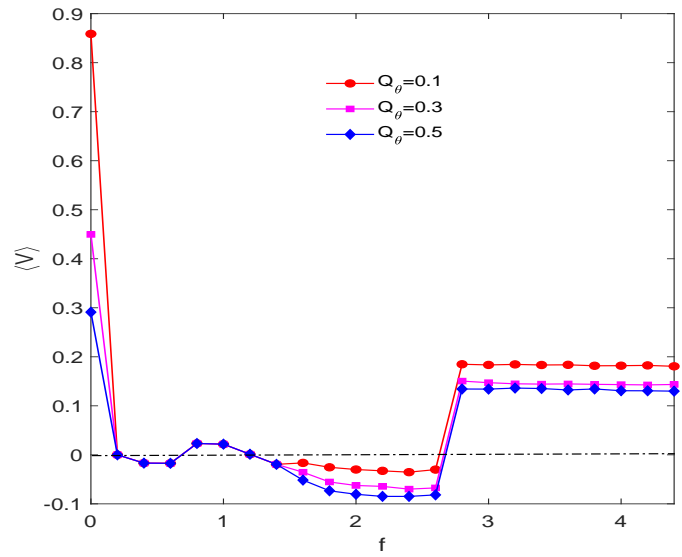}
\caption{The average velocity $\langle V\rangle$ as a function of $f$ with different $Q_{\theta}$. The other parameters are $Q_x=Q_y=2.0$, $v_0=2.0$, $\omega=0.2$, $\Delta=0.2$, $a=1/(2\pi)$, $b=1.2a$, $L=1.0$, $\mu=1.0$, $\tau_x=\tau_y=1.0$.}
\label{Vf}}
\end{figure}
Figure \ref{Vf} shows the dependence of $\langle V\rangle$ on the time periodic force amplitude $f$  with different $Q_{\theta}$. We find $\langle V\rangle$ exhibits complicated behavior with increasing force amplitude $f$. $\langle V\rangle$ is positive when $f=0$, so the particles move in $+x$ direction when the time periodic force is not exist. With $f$ increasing, the moving direction rapidly changes to $-x$ direction($\langle V\rangle <0$) when $f\approx0.2$, and changes to $+x$ direction when $f\approx0.7$, and changes to $-x$ direction again when $f\approx1.2$, and changes to $+x$ direction again when $f\approx2.7$. In the end, $\langle V\rangle-f$ nearly parallel to $f$ axis and do not change.

\begin{figure}
\center{
\includegraphics[height=8cm,width=10cm]{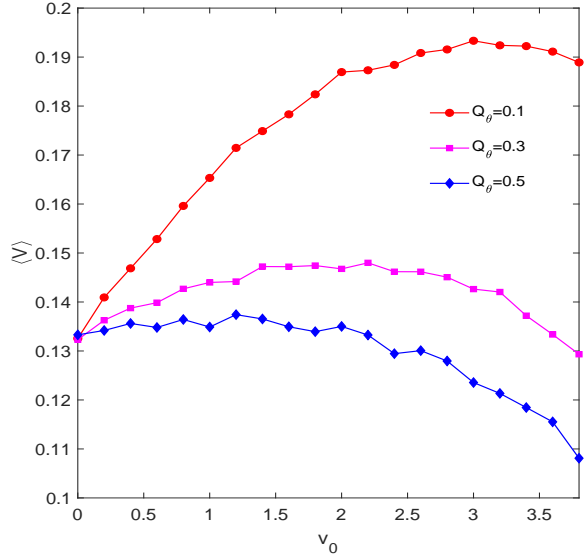}
\caption{The average velocity $\langle V\rangle$ as a function of $v_0$ with different $Q_{\theta}$. The other parameters are $Q_x=Q_y=2.0$, $\omega=0.2$, $\Delta=0.2$, $a=1/(2\pi)$, $b=1.2a$, $L=1.0$, $f=3.0$, $\mu=1.0$, $\tau_x=\tau_y=1.0$.}
\label{Vv0}}
\end{figure}
Figure \ref{Vv0} shows the dependence of $\langle V\rangle$ on the self-propelled speed $v_0$ with different$Q_{\theta}$. As shown, we find $V$ has a maximum with increasing $v_0$. So proper value of self-propelled speed should better for particles directional movement, but too large $v_0$ should  restrain this phenomenon.

\begin{figure}
\center{
\includegraphics[height=8cm,width=10cm]{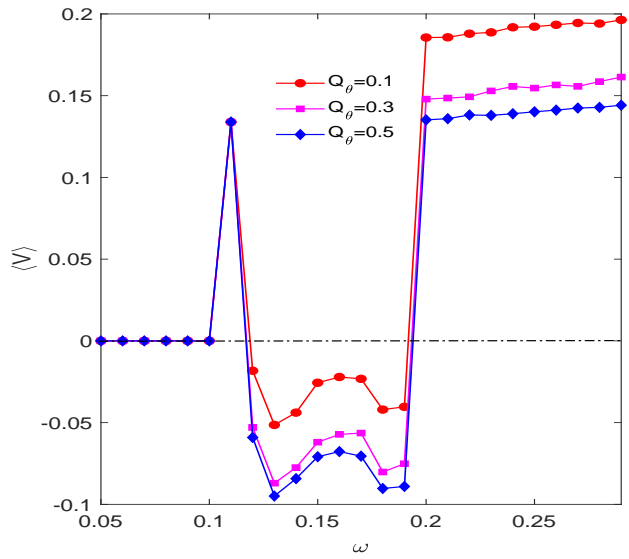}
\caption{The average velocity $\langle V\rangle$ as a function of $\omega$ with different $Q_{\theta}$. The other parameters are $Q_x=Q_y=2.0$, $v_0=2.0$, $\Delta=0.2$, $a=1/(2\pi)$, $b=1.2a$, $L=1.0$, $f=3.0$, $\mu=1.0$, $\tau_x=\tau_y=1.0$.}
\label{Vomega}}
\end{figure}
Figure \ref{Vomega} shows the dependence of $\langle V\rangle$ on the periodic force angular frequency $\omega$. We find the directional movement should disappear($\langle V\rangle\approx0$) when the angular frequency $\omega\leq0.1$. The particles move in $+x$ direction when $\omega=0.11$, and move in $-x$ direction when $0.12\leq\omega\leq0.19$. When $\omega \geq0.20$, the particles moving direction changes to $+x$ again, and the direction movement speed $\langle V\rangle$ increase with increasing $\omega$.

\begin{figure}
\center{
\includegraphics[height=8cm,width=10cm]{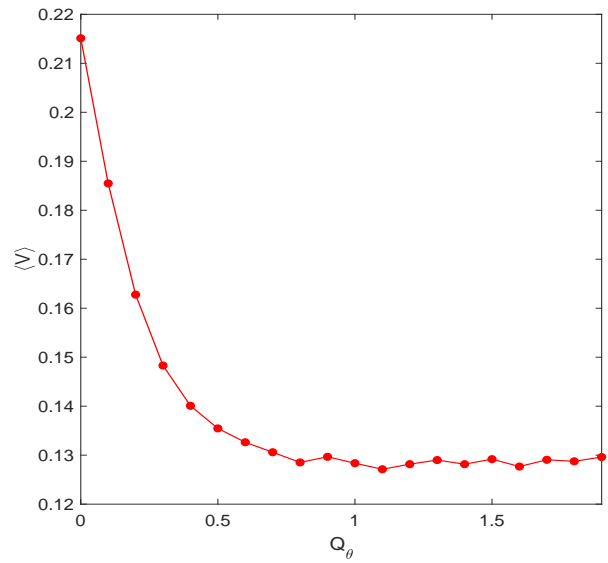}
\caption{The average velocity $\langle V\rangle$ as a function of $Q_{\theta}$. The other parameters are $Q_x=Q_y=2.0$, $v_0=2.0$, $\omega=0.2$, $\Delta=0.2$, $a=1/(2\pi)$, $b=1.2a$, $L=1.0$, $f=3.0$, $\mu=1.0$, $\tau_x=\tau_y=1.0$.}
\label{VQtheta}}
\end{figure}
The average velocity $\langle V\rangle$ as a function of self-propelled angle noise intensity $Q_{\theta}$ is shown in Fig.(\ref{VQtheta}). It is obviously that $\langle V\rangle$ decreases with increasing $Q_{\theta}$. Small value of $Q_{\theta}$ helps to the particles directional movement, but large  $Q_{\theta}$ should restrain this phenomenon. This result is just consistent with the results of Figs.(\ref{VQx}) and (\ref{VQy}).

\section{\label{label4}Conclusions}
In this paper, we numerically studied the transport phenomenon of self-propelled particles confined in a two-dimensional smooth channel with colored noise. We find large $x$ direction noise intensity should reduce particles directional movement when the angle Gaussian noise intensity is small. When $Q_{\theta}$ is large, the directional movement speed has a maximum with increasing $Q_x$. $y$ directional noise has negligible impact on the $x$ directional movement. The movement direction changes more than once with increasing periodic force amplitude. Large angle Gaussian noise intensity should  weaken the particles directional transport.
\section{Acknowledgments}
Project supported by Natural Science Foundation of Anhui Province(Grant No:1408085QA11) and the National Natural Science Foundation of China (No:11404005).

\bibliographystyle{elsarticle-num}

\end{document}